\documentclass[prl,aps,amsmath,amsfonts,twocolumn]{revtex4-2}
\usepackage[pdftex]{graphicx} 
\usepackage{chapterbib} 
\begin{document}
\newcommand{\asinh}{\mathrm{asinh}} 
\newcommand{\hc}{\mathrm{h.c.}}
\title{Hall response in interacting bosonic and fermionic ladders}

\date{\today}
\author{R. Citro}
\affiliation{Dipartimento di Fisica ”E.R. Caianiello”, Universit\`{a} degli Studi di Salerno and CNR-SPIN c/o University of Salerno,Via  Giovanni  Paolo  II,  132,  I-84084  Fisciano  (Sa),  Italy}
\affiliation{INFN,  Sezione  di  Napoli,  Gruppo  collegato  di  Salerno,  I-84084  Fisciano  (SA),  Italy}
\author{T. Giamarchi}
\affiliation{DQMP,  University  of  Geneva,  24  Quai  Ernest-Ansermet,  CH-1211  Geneva,  Switzerland} 
\author{E. Orignac}
\affiliation{Univ Lyon, Ens de Lyon,  CNRS, Laboratoire de Physique, F-69342 Lyon, France}
\begin{abstract}
  We use bosonization, retaining band curvature terms, to analyze the Hall response of interacting bosonic and fermionic two-leg ladders threaded by a flux. 
  We derive an explicit expression of the Hall imbalance in a perturbative expansion in the band curvature, retaining fully the interactions. We show that the flux dependence of the Hall imbalance allows to distinguish the two phases (Meissner and Vortex) that are present for a bosonic ladder. For small magnetic field we relate the Hall resistance, both 
  for bosonic and fermionic ladders, to the density dependence of the charge sitffness of the system in absence of flux. Our expression unveil a universal interaction-independent behavior in the Galilean invariant case.
\end{abstract}
\maketitle
\begin{cbunit}
Since its discovery in 1879 \cite{hallnew} the Hall effect has become a remarkable tool for studying solid-state systems. For a single band, the sign of the Hall coefficient $R_H$, defined as the ratio of the induced electric field to the product of the current density and the applied magnetic field, permits the extraction of the effective charge $q$ and carrier density $n$, as $R_H\sim -1/nq$ in conventional conductors, indicating whether the carriers are electrons or holes \cite{ziman_solid_book}. The Hall effect and its quantum version have thus naturally found applications in metrology for sensitive measurements of magnetic fields and for resistance
standards \cite{klitzing1980}. In non-interacting systems,  the Hall effect is interpreted as a manifestation of the topological
properties of quantum states, such as the Berry curvature in the anomalous Hall systems \cite{review_niu_2010}, and topological invariants in the Integer Quantum Hall effect \cite{thouless_1982}. Studies of the dynamical version of the Hall response have also become relevant in fields addressing topological quantum transport \cite{citro_2023} and synthetic dimensions \cite{celi_synthetic_dimensions_cold}. 

However, the understanding of Hall effect in presence of interactions still remains a fundamental theoretical challenge. With strong magnetic field, the fractional quantum hall effect \cite{tsui_1982} has revealed excitations with fractional charge and anyonic statistics \cite{laughlin_1983}. In the opposite limit of small magnetic field, a complete interpretation of the Hall effect is still missing and few studies are present \cite{leon_hall,zotos_2000,auerbach_2018}. Recently a numerical study of the Hall coefficient in a quasi-one dimensional system, has predicted a universal behavior for the Hall coefficient above an interaction threshold \cite{greschner_universal_2019}. In the case of $N-$leg ladder systems with $SU(N)$ symmetry, the Hall imbalance, i.e. the difference of particle number between upper and lower legs, was shown to take the classical value  $R_H=1/n$ \cite{greschner_universal_2019}, a prediction confirmed in a quantum simulation with strongly interacting ultracold fermions \cite{zhou_2022}. The possibility to reliably measure the Hall effect in strongly correlated ladders thus prompts for an analytical calculation of the Hall polarization and voltage that can shed light on the many-body effect controlling such quantities in a correlated system. Such endeavour has proven however elusive for the moment since the most common approximations 
done to deal with interacting one dimensional systems lead to an artificial particle-hole symmetry and thus to zero Hall effect.

In this Letter we focus, for a two-leg ladder of bosons or fermions threaded by a magnetic flux, on the study of the Hall effect.  We use a bosonization approach  \cite{giamarchi_book_1d}, but properly taking into account band curvature \cite{Leon-PRB-2007,lopatin_q1d_magnetooptical}. We obtain analytically both the Hall imbalance and Hall voltage and show that the latter is 
given by the derivative with respect to density of the logarithm of the charge stiffness of the system without flux. 
For a Galilean invariant case, this relation leads 
back to the standard formula \cite{ziman_solid_book} related to the carrier density, while in the more general case this remarkable connection between two important 
transport coefficients encodes the many-body effects triggered by interactions in the Hall response. 
Beyond the clear potential of our formula to measure Hall voltages and clarify the exotic Hall response of strongly correlated solid-state conductors, 
our work paves the way to the investigation of the topological transport properties of strongly correlated systems of matter.

We study the ladder shown in Fig.~\ref{fig:ladder} and populated by bosons or fermions. 
\begin{figure}
\centering\includegraphics[width=\columnwidth]{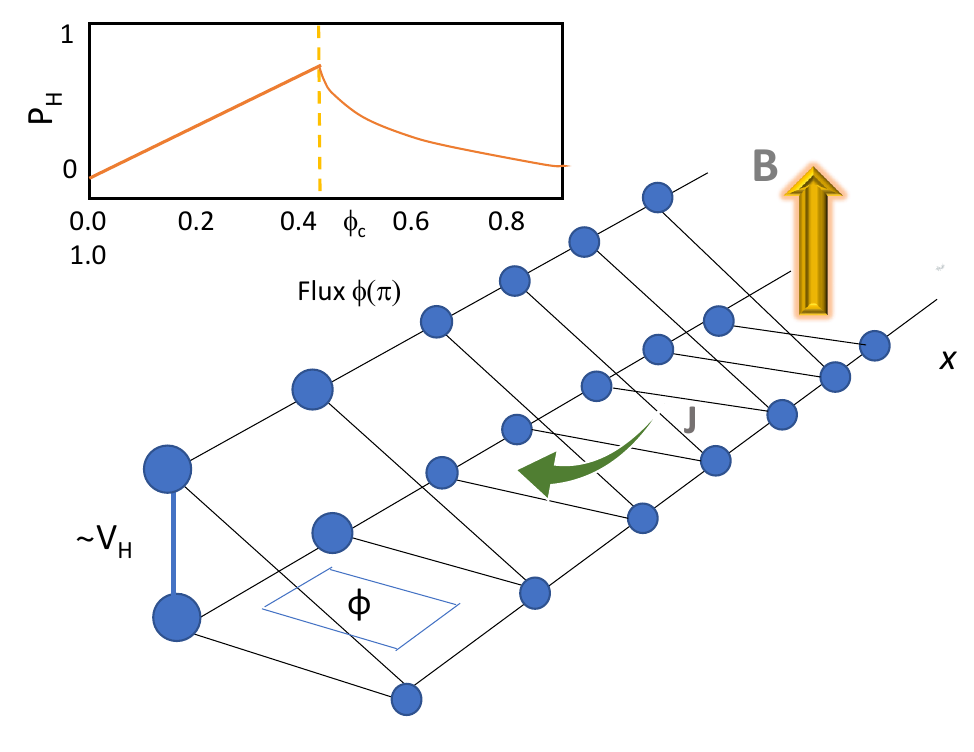}
    \caption{\label{fig:ladder}  Two-leg ladder (in real or synthetic dimensions), with an applied magnetic field perpendicular to the plane and current along the $x$ direction. Due to Hall effect the current is deflected and a Hall voltage can be measured along the rungs. $\phi$ is the flux of the magnetic field in the plaquette. Although the figure shows also a lattice along the 
    chain direction, the chains can also be uniform. The linear ramp scheme for the measure of the Hall voltage is also shown.The inset shows the qualitative trend of the Hall polarization as a function of the applied flux.}
\end{figure}
Such a system can be realized in platforms ranging from cold atoms, both with synthetic dimensions \cite{fallani_biperiodic_cold} or on a lattice \cite{atala_2014,stuhl_ladder_2015,chalopin_hall_2020} to condensed matter such as graphene nanoribbons \cite{wang_nanoribbons_2021,lou_nanoribbons_2021} or Josephson junction arrays \cite{vanoudenaarden_josephson_mott,haviland_josephson1d,Fazio-PREP-2001,duty_bose_glass_1d_josephson,Kuzmin2019}.

Let us first examine bosonic systems. Without magnetic field, the generic ladder model has the Hamiltonian:
\begin{equation} \label{eq:genericbos}
 H = H_1 + H_2 - t_\perp \int dx \, (b^\dagger_1(x) b_2(x) + \hc).
\end{equation}
With a lattice along the chains, the integral along $x$ in (\ref{eq:genericbos}) becomes a discrete sum over the sites $j$. 
$H_{1,2}$ describe interacting bosons along the chains, while $t_\perp$ is the tunneling along the rung. Typical realizations of such systems for cold atoms involve 
contact interactions and are the Lieb-Liniger \cite{lieb_bosons_1D} model for continuous chains \cite{bouchoule_atom_2011,fabbri_dynamical_2015,meinert_probing_2015} or the Bose-Hubbard model \cite{jaksch_bose_hubbard} for chains with a lattice \cite{koehl_1D_bose_gases_opt_lat,hofferberth2007,haller_mott_1d}. When the ladder is realized using synthetic dimensions \cite{celi_synthetic_dimensions_cold,Barbarino2016}, in addition to 
the interactions along the chains an interaction across the rung also exists.

Not to be tied to a particular microscopic model, we use the fact that they all have the low energy physics of a Tomonaga-Luttinger liquid (TLL)
\cite{haldane_bosons,glazman_josephson_1d,basko_josephson1d,wu_josephson_1d,giamarchi_book_1d,cazalilla_review_bosons}. Analyzing the TLL model for the 
bosonic version of Fig.~\ref{fig:ladder} thus describes all the microscopic models in a unified  way. 
The single chain Hamitonian is ($p=1,2$ and we set $\hbar=1$):
\begin{multline}\label{eq:TLLdisp}
 H_p = \int \frac{dx}{2\pi} \left[u K (\pi \Pi_p)^2 +\frac u K (\partial_x \phi_p)^2\right] \\  
 + \int dx \left[\alpha (\partial_x \phi_p)^3 +  \gamma (\pi  \Pi_p)^2 \partial_x \phi_p\right].
\end{multline}
The fields 
$\partial_x\theta_p = \pi \Pi_p$ and $\phi_p$ represent respectively the collective long wavelength excitations in the $p$-th chain  associated to the gradient of the superfluid phase $\theta_p$ and to the deviations of the density $\rho_p(x)$ of the bosons $\rho_p(x)-\rho_0 \simeq -\frac1\pi \partial_x \phi_p(x)$ from the 
average density $\rho_0$. They obey the commutation relation $\lbrack \phi_p(x),\pi \Pi_{p'}(x')\rbrack=i\pi \delta{_pp'} \delta (x-x')$. In the first line, the sound velocity $u$ and the dimensionless TLL parameter $K$  can be computed for any given microscopic 
Hamiltonian \cite{giamarchi_book_1d,cazalilla_review_bosons,amico04_boson_integrable_review,kuehnerMonien1998,Ejima-EPL-num,ejima_tl_mott,Crepin2011,pandey_ladder}. The parameter $K$ controls the algebraic decay of the correlation functions \cite{giamarchi_book_1d}. 
The second line corresponds to
irrelevant operators that must be retained for the Hall problem. They represent the terms generated by band curvature 
\cite{schick_flux_1968,haldane_bosonisation,giamarchi_curvature} and break particle-hole symmetry. 
This last point is crucial for a proper description of the 
Hall effect since particle-hole symmetry immediately 
implies a zero Hall voltage. The parameters $\alpha,\gamma$ are given by \cite{matveev_equilibration_2013,pereira_dynamical_2007,matveev_effective_2016}
\begin{equation}
   \alpha = -\frac{\partial}{\partial \rho_0} \left(\frac{u}{6\pi^2 K}\right) \quad;\quad 
   \gamma = -\frac{\partial}{\partial \rho_0} \left(\frac{uK}{2 \pi^2}\right). \label{eq:gamma-phen} 
\end{equation}
In a galilean invariant system \cite{haldane_bosons}, (\ref{eq:gamma-phen})  simplifies to $\gamma = -1/(2\pi m)$, with $m$ the particle mass. Furthermore, in the case of hardcore bosons or spinless fermions with a quadratic dispersion, $\alpha = -1/(6\pi m)$
\cite{schick_flux_1968,haldane_bosonisation}. 
The generalization to the case with interchain interactions is discussed in the Supplemental material (SM) \cite{supp}.

In addition to the Hamiltonians of the two independent chains 
(\ref{eq:TLLdisp}) the  boson tunnelling \cite{orignac_2chain_bosonic,orignac01_meissner} between the chains gives 
\begin{equation}\label{eq:TLLtun}
  H_{\mathrm{tunn.}} = -\frac{t_\perp}{\pi a} \int dx \cos (\theta_1(x) -\theta_2(x)).  
\end{equation}
With a magnetic field and in the Landau gauge, the vector potential along chain $p$ is $A_p=(-1)^{p-1} \frac{Ba} 2$, with $a$ the interchain distance. The vector potential is zero along the rung.  As shown in the SM \cite{supp} this leads to the replacement  $\partial_x \theta_p \to \partial_x\theta_p - e A_p$ in (\ref{eq:TLLdisp}). The longitudinal current in chain $p$ is obtained from the gauge field dependence of the Hamiltonian $j^x_p(x) = -\frac{\delta H_p}{\delta A_p(x)}$. We define the Hall polarization \cite{greschner_universal_2019}:
\begin{equation}\label{eq:hall-polarization}
  P_H= N_1-N_2=\int dx [\rho_1(x) -\rho_2(x)] , 
\end{equation}
which is easily detected \cite{Aidelsburger2013,stuhl_ladder_2015,mancini_ladder_synthetic}. We introduce the symmetric and antisymmetric basis \cite{orignac98_vortices,orignac01_meissner} $\phi_{\rho,\sigma} =\frac {\phi_1\pm\phi_2}{\sqrt{2}}$; $\Pi_{\rho,\sigma} = \frac {\Pi_1\pm \Pi_2}{\sqrt{2}}$
and the corresponding vector potentials 
$A_{\rho,\sigma}=\frac 1 {\sqrt{2}} (A_1\pm A_2)$,
yielding the full Hamiltonian 
$H = \sum_{\nu=\rho,\sigma} H_\nu + H_{\text{hop.}} + H_{\text{curv.}}$ with
\begin{equation} \label{eq:boso-asym}
  H_\nu = \int \frac{dx}{2\pi} \left[ u_\nu K_\nu (\pi \Pi_\nu - e A_\nu)^2 +\frac {u_\nu} {K_\nu} (\partial_x \phi_\nu)^2\right], 
\end{equation}
with the tunneling term $H_{\text{hop.}} = -\int dx \frac{2 t_\perp}{a_0} \cos(\sqrt{2} \theta_\sigma)$ and the band curvature term 
\begin{multline}\label{eq:saterm}
 H_{\text{curv.}} = \int \frac{dx}{\sqrt{2}} \left[\gamma \partial_x \phi_\rho (\pi \Pi_\rho -e A_\rho)^2 + \alpha (\partial_x \phi_\rho)^3\right] \\
 +\left\{ \partial_x \phi_\rho \left[ \gamma (\pi \Pi_\sigma - e A_\sigma)^2 + 3\alpha (\partial_x \phi_\sigma)^2  \right] \right. \\ 
 + \left. 2\gamma (\pi \Pi_\sigma - e A_\sigma)  (\pi \Pi_\rho - e A_\rho) \partial_x \phi_\sigma       \right\},    
\end{multline}
The parameters are $A_\rho=0$, $A_\sigma = \frac{Ba}{\sqrt{2}}$, $u_\nu=u$, $K_\nu=K$. The currents and the Hall polarization become 
\begin{equation}
\begin{split}
 j^x_{\rho,\sigma}(x) &= (j^x_1\pm j^x_2)(x) = -\sqrt{2} \frac{\delta H}{\delta A_{\rho,\sigma}(x)}, \\
  P_H &= -\frac{\sqrt{2}}{\pi} \int_{-\infty}^{+\infty} \partial_x \phi_\sigma.
\end{split}
\end{equation}

Since the Hall effect vanishes with particle-hole symmetry,
it is \emph{perturbative} in the band curvature \cite{leon_hall,lopatin_hall_luttinger}, in agreement 
with free fermions calculations. We use this property to build a \emph{perturbative} approach 
of the Hall imbalance and voltage in the band curvature while 
retaining the \emph{full non-perturbative} description in the interactions.
This allows for the first time for an interacting 
system to explicit compute the the Hall effect in a regime of weak interchain tunneling.
In this framework the equilbrium imbalance is 
\begin{equation}\label{eq:hallaverage}
  \frac{\langle P_H\rangle}{L} =  - \frac{\left\langle T_\tau e^{-\int_0^\beta d\tau H_{curv.}(\tau)} \frac{\sqrt{2} \partial_x \phi_\sigma(x,0)}{\pi} \right\rangle_{H_\rho+H_\sigma+H_{\text{hop.}}}}{ \left\langle T_\tau e^{-\int_0^\beta d\tau H_{curv.}(\tau)}\right\rangle_{H_\rho+H_\sigma+H_{\text{hop.}}}},  
\end{equation}
We expand the exponentials in (\ref{eq:hallaverage}) to first order in $\alpha,\gamma$, leading 
naturally to an expression of $\langle P_H\rangle$  at first order in the band 
curvature terms $\gamma,\alpha$ since the zero order term is cancelled by particle-hole symmetry. 

Only the last line of (\ref{eq:saterm}) contributes. It couples the local density imbalance between the chains $\propto \partial_x \phi_\sigma$  to the difference of current between the chains $\propto (\pi \Pi_\sigma - e A_\sigma)$ and to the total longitudinal current $\propto (\pi \Pi_\rho  -e A_\rho)$. Quite remarkably, although the longitudinal total current passing through the system can be a complicated quantity as a function of the gauge potential (flux), the expectation value of the current simply factorizes. Since we ultimately divide by it, it is 
not necessary to explicitly compute it, leading to further simplifications in our formula since only the antisymmetric correlator is  needed.
The Hall polarization reads 
\begin{multline} \label{eq:hallbosfinal}
    \langle P_H\rangle \simeq  \frac {\sqrt{2}\gamma \langle j_\rho \rangle L} {e u_\rho K_\rho} \int_0^L dy \int_0^\beta d\tau  \\
    \left\langle T_\tau \left[(\pi \Pi_\sigma - e A_\sigma) \partial_x \phi_\sigma \right]_{(y,\tau)} 
     \partial_x \phi_\sigma(0,0) \right\rangle_{H_\sigma+H_{\text{hop.}}}. 
\end{multline}

(\ref{eq:hallbosfinal}) is one of the central results of the paper. It provides an explicit formula for an analytical or numerical calculation of the Hall imbalance once normalized with the longitudinal current. To finish the calculation we separate the two possible regimes of 
such a two-leg ladder \cite{orignac01_meissner,cha2011,tokuno2014,piraud2014b,didio2015}: a Meissner phase at low magnetic field in which currents circulate only 
along the legs ($\langle \Pi_\sigma \rangle =0$) and,  
at higher fields, a vortex phase \cite{orignac01_meissner} where currents also circulate along the rungs ($\langle \Pi_\sigma \rangle  \ne 0$). The two phases are separated by a commensurate-incommensurate transition \cite{japaridze_cic_transition,pokrovsky_talapov_prl,schulz_cic2d,Citro2018,Orignac2017}.

We first analyze the Meissner regime, for which we can take the limit $B \to 0$.
In such phase, the symmetry $\phi_\sigma \to -\phi_\sigma$ and $\Pi_\sigma \to - \Pi_\sigma$, implies $\langle \Pi_\sigma \partial_x \phi_\sigma \partial_x \phi_\sigma \rangle=0$, and the imbalance reduces to (see SM~\cite{supp})
\begin{equation}
  \langle P_H\rangle \simeq
    \frac{\pi^2 B a \gamma \langle j_\rho \rangle}{2 u_\rho K_\rho} \chi_{\sigma \sigma}(q=0,\omega_n=0).   
\end{equation}
Thanks to (\ref{eq:saterm}), the first order perturbation in the curvature gives back the expected
proportionality both in the total current and in the applied flux. 
The susceptibility $\chi_{\sigma \sigma}$ measures the charge imbalance in response to a chemical potential difference between the two chains.
Within linear response theory \cite{supp}, $\chi_{\sigma \sigma} = \frac{2 K_\sigma}{ \pi u_\sigma}$, leading to
\begin{equation}\label{eq:imbalance-boso-perturb}
   \frac{\langle P_H\rangle}{B L \langle j_\rho \rangle} \simeq \frac{a \pi \gamma K_\sigma }{u_\rho K_\rho u_\sigma} .
\end{equation}
giving explicitly the Hall imbalance, normalized with the longitudinal current and field, in term of the TLL parameters of the system.

Beyond a certain magnetic field, 
in the vortex phase, the symmetry becomes $\phi_\sigma \to -\phi_\sigma$ and $\Pi_\sigma \to 2\langle \Pi_\sigma \rangle -\Pi_\sigma$, resulting in  $\langle (\Pi_\sigma -\langle \Pi_\sigma \rangle)  \partial_x \phi_\sigma \partial_x \phi_\sigma \rangle=0$. As a result
\begin{equation}
  \frac{\langle P_H \rangle}{L} \simeq\frac{\sqrt{2} \gamma \langle j_\rho\rangle K_\sigma}{e u_\rho u_\sigma K_\rho} (\pi \langle \Pi_\sigma \rangle -e A_\sigma).    
\end{equation}
The contribution of $\langle \Pi_\sigma\rangle \ne 0$ lowers the Hall imbalance compared with the Meissner phase. As $B$ increases, the expectation value $(\pi \langle \Pi_\sigma \rangle -e A_\sigma) \to 0$  \cite{orignac01_meissner}, so the Hall polarization tends to zero a signature of the decoupling of the chains at high magnetic field. 

We now turn to the calculation of the Hall \emph{voltage}, obtained by adding a potential difference $V_H$ between the two chains \cite{prelovsek_hall_1999,buser_hall_ladder}, 
$-\frac{eV_H}{\pi\sqrt{2}} \int dx \partial_x \phi_\sigma$, such that $\langle P_H \rangle =0$. 
In the Meissner phase, the Hall voltage is 
\begin{equation}\label{eq:hall-voltage-bosons}
  V_H=-\frac{\pi^2 \gamma \langle j_\rho \rangle}{e u_\rho K_\rho} B a = - \langle j_\rho \rangle \frac{B a}{2e} \frac 1 {u_\rho K_\rho} \frac{\partial}{\partial \rho_0} (u_\rho K_\rho), 
\end{equation}
where we have used the relation~(\ref{eq:gamma-phen}). 
Since the longitudinal current density $I_x=\langle j_\rho\rangle/2$ this reduces to 
\begin{equation}\label{eq:hall-resistance-bosons}
 R_H=\frac{a}{e} \frac{\partial}{\partial \rho_0} [\ln (u_\rho K_\rho)]. 
\end{equation}
This formula, which is the second central result of the paper, quite remarkably relates the Hall voltage to the charge stiffness $\mathbb{D}$ \cite{kohn_stiffness} (weight of the Drude peak in conductivity, or response of the ground state energy of the systems to a twist in boundary conditions) in the absence of flux. Indeed the charge stiffness is given for a TLL by $u_\rho K_\rho$ \cite{giamarchi_book_1d}. Although we have derived (\ref{eq:hall-resistance-bosons}) in the absence of interchain interactions we show in the SM~\cite{supp} that this formula remains valid even in their presence.

Several immediate consequences can be extracted from (\ref{eq:hall-resistance-bosons}). In a Galilean invariant model \cite{haldane_bosons}, $uK = \frac{\pi \rho_0}{m}$,  and the Hall resistance reduces to 
\begin{equation}\label{eq:rh_naive}
  R_H=\frac{a B}{e \rho_0},  
\end{equation}
One thus recovers the ``naive'' \cite{ziman_solid_book} expression of the non-interacting Hall effect (for fermionic particles) for which the Hall effect simply measures the inverse carrier density. This shows analytically for the first time that these results extend to interacting particles and bosonic
statistics, provided Galilean invariance holds. Although a general argument leading to (\ref{eq:rh_naive}) in two dimensions was known, it required \emph{both} Lorentz invariance and translational invariance \cite{girvin_houches}. In the case of an artificial gauge field, the first condition may not be satisfied. More importantly, translational invariance is only satisfied along the chains, preventing the conservation of transverse momentum. In the SM \cite{supp}, we give a simple argument applicable for an artificial gauge fields and hopping in the transverse direction to show that in the Galilean invariant case, $R_H\sim 1/\rho_0$ holds beyond perturbation theory. 

Formula (\ref{eq:hall-resistance-bosons}) thus paves the way to efficiently use the Hall voltage to identify the phase transitions occuring in an interacting system. For example, it allows to distinguish the Meissner phase, where the Hall resistance remains constant and related to the inverse density, and the vortex phase, where the Hall resistance decreases as a function of the magnetic field. (\ref{eq:hall-resistance-bosons}) also allows to compute, in the absence of Galilean invariance, the impact of the interactions on the Hall voltage. As we show in the SM~\cite{supp} this formula is compatible with numerical calculations of the Hall imbalance in the Meissner phase \cite{buser_hall_ladder} for the Bose-Hubbard ladder.

In particular (\ref{eq:hall-resistance-bosons}) shows that 
the Hall voltage will signal phases in which $\mathbb{D}$ is anomalous such as e.g. Mott insulating phases. Close a commensurate filling, umklapp operators  
\cite{giamarchi_mott_shortrev,giamarchi_book_1d} lead to a renormalization of $u_\rho K_\rho$. If the umklapp is irrelevant or away from commensurability we can simply use 
in (\ref{eq:hall-resistance-bosons}) the fixed point value $u_\rho^* K_\rho^*$. When the umklapp is relevant, the ladder becomes insulating, $\langle j_\rho \rangle=0$, 
and obviously $\langle P_H \rangle=0$. Moving away from commensurability restores the TLL \cite{giamarchi_mott_shortrev} via a commensurate-incommensurate transition 
\cite{japaridze_cic_transition,pokrovsky_talapov_prl,schulz_cic2d}. In the vicinity of the transition \cite{schulz_cic2d}, the renormalized TLL exponent goes to  a constant $K_\rho^*=1/2$, 
while the velocity vanishes as $u_\rho \sim |\rho_0-\rho_{0,c}|$. As a result, (\ref{eq:hall-resistance-bosons}) predicts a \emph{divergent} Hall resistance 
\begin{equation}
 R_H \sim \frac{a}{e} \frac{\mathrm{sign}(\rho_0-\rho_{0,c})}{|\rho_0-\rho_{0,c}|},    
\end{equation}
with a sign reflecting the nature of the effective carriers. A similar divergence of the Hall resistivity is also obtained in the case of a nearly empty band, where  $K_\rho \to 1$ while the velocity $u_\rho\sim \rho_0$ and in the case of a nearly filled band of hard core bosons, where $u_\rho \to 1-\rho_0$. These predictions can be directly tested either in numerical studies or potentially quantum simulations \cite{Crepin2011}. More properties of the Hall voltage are discussed in the SM~\cite{supp}.

Let us finally obtain similar results for the fermionic case, also be realizable in cold atoms\cite{sompet_2022,hirthe_2023,zhou_2022,pagano_ytterbium_cold,Salomon2019} or condensed matter \cite{wang_nanoribbons_2021,lou_nanoribbons_2021}
experiments.
Without interchain hopping, a spin-1/2 fermions chain with repulsive contact interaction is described in the continuum by the Gaudin-Yang model \cite{gaudin_fermions,yang_fermions} and on a lattice by the Hubbard model \cite{lieb_hubbard_exact}.  
Due to the $\text{SU(2)}$ symmetric interactions, the Zeeman field leaves the spin modes \cite{schulz_su2} gapless in contrast with the generic spinless fermion ladder \cite{nersesyan_2ch,giamarchi_spin_flop}. The Hamiltonian for fermions with a quadratic dispersion along the chains reads
\begin{multline} \label{eq:hamiltonian}
  H = \sum_{p=1,2} \int dx \frac{\psi^\dagger_p (x)\left(\frac \hbar i \nabla - e A_p\right)^2 \psi_p(x)}{2m}-\mu \psi^\dagger_p(x) \psi_p(x)  \\  
      -t_\perp \int dx (\psi^\dagger_1 \psi_2 + \psi^\dagger_2 \psi_1)  + U \int dx \rho_1(x) \rho_2(x)  \\ 
     + \frac 1 2 \int dx V(x-x') [\rho_1(x)\rho_1(x')+\rho_2(x)\rho_2(x')] 
\end{multline}
where $\psi_p$ annihilates a fermion in the chain $p$, $\rho_p =\psi^\dagger_p \psi_p$,  $A_p$ is the vector potential in the Landau gauge along the chain $p$, $\mu$ the chemical potential, $m$ the mass, $e$ the fermion charge, $V(x)$ the short ranged-intrachain interaction, $U$ the interchain interaction and $t_\perp$ the interchain hopping.
Linearizing the spectrum around the Fermi points, with $\psi_p(x)=e^{i k_F x} \psi_{Rp}(x)+e^{-i k_F x} \psi_{Lp}(x)$, with $k_F=\pi \rho_0$ the Fermi wavevector and $v_F=k_F/m$ the Fermi velocity, we bosonize using standard formulas \cite{nersesyan_2ch,orignac97_ladder} (see SM \cite{supp}) in terms of symmetric $\phi_\rho; \theta_\rho$ and antisymmetric $\phi_\sigma; \theta_\sigma$ fields.
The Hall polarization has the bosonized expression
\begin{equation} \label{eq:imbalance-bosonized}
  P_H=2\int \frac{dx}{\pi a_0} \cos \sqrt{2} \theta_\sigma \cos \sqrt{2} \phi_\sigma.  
\end{equation}
Without band curvature and magnetic flux, the ground state of the model \cite{nersesyan_2ch,orignac97_ladder} is well known. The field $\theta_\sigma$ is ordered 
with an expectation value $\langle \theta_\sigma \rangle = \pi/\sqrt{8}$, while the field $\theta_\rho$ remains gapless. The curvature terms yield a contribution to the Hamiltonian 
\begin{equation}
-\frac{eA_\sigma}{\pi m\alpha} \int dx (\sqrt{2} \pi \Pi_\rho - e A_\rho) \cos \sqrt{2} \theta_\sigma \cos \sqrt{2} \phi_\sigma,  
\end{equation}
which results in an expression of $\langle P_H \rangle$ proportional to $B \langle j_\rho \rangle \chi_{\perp}$, with $\chi_{perp}$
the in-plane magnetic susceptibility of an easy axis antiferromagnetic XXZ spin-1/2 chain at wavevector $t_\perp/v_F$.  Since the XXZ chain is close to the isotropic point, for large enough $t_\perp/v_F$, the in-plane susceptibility is the same as in the isotropic chain. Up to logarithmic corrections \cite{schulz_su2}, $\chi_\perp \sim 1/v_F$, and the perturbative behavior of the imbalance in the fermionic ladder is the same as in the bosonic ladder. 

To find the Hall resistance, we follow the same path than for bosons. In the bonding-antibonding basis, after bosonization, the potential difference reads
\begin{eqnarray}
 V_\perp = -\frac{e V_H}{\pi a_0} \int dx \cos \sqrt{2} \theta_\sigma \cos \sqrt{2} \phi_\sigma,  
\end{eqnarray}
and we have to determine $V_H$ such that  $\langle P_H \rangle_{H+V_\perp}=0$. To lowest order in $1/m$, we obtain
the same expression for $R_H$ as in the bosonic case, (\ref{eq:hall-resistance-bosons}). We have explicitly checked in the SM \cite{supp} that at low flux, a relation similar 
to (\ref{eq:hall-resistance-bosons}) is valid for a two-leg ladder of non-interacting fermions. 

In conclusion, we have derived an analytic expression for the Hall imbalance and Hall voltage for a two-leg ladder threaded by a flux using a bosonization approach retaining band curvature terms. One of our central results relates the Hall resistance to the logarithmic derivative of the charge stiffness with respect to the density. This formula, applicable both for 
bosons and fermions -- as long as there is only one gapless mode -- allows a direct and practical calculation of the interaction effects on the Hall voltage. In particular it shows that 
the Hall resistance  diverges close to a Mott insulating phase, on the other side, in the Galilean invariant case, it depends only on density as in the non-interacting case. 

The Hall response studied here can be detected in the controllable quantum simulator of a two-leg
ladder with interacting bosonic \cite{atala_2014,stuhl_ladder_2015,chalopin_hall_2020} and fermionic \cite{livi_clock_2016,han_ladder_2022} atoms. A first measure of Hall response in a quantum simulator with strongly interacting ultracold fermions has recently appeared \cite{zhou_2022}, opening the way towards further experimental study in different regimes of the interaction and magnetic field. For bosons instead  a systematic measurement of the Hall coefficient was
shown recently in Ref. \cite{genkina_2015} and our study is compatible with the experimental platforms of Refs.~\cite{aidelsburger_2015,genkina_2019,zhou_2022}. Another experimental platform could be Josephson junctions networks \cite{vanoudenaarden_josephson_mott,vanoudenaarden_josephson_localization,fazio_josephson_junction_review,duty_bose_glass_1d_josephson,Kuzmin2019}, where the Hall resistivity could be directly measured. A last possible platform, transmon qubits~\cite{ye2019qubit,xiang2022simulating} have recently been shown to realize the hardcore boson ladder, and there are recent proposals~\cite{guan2020} to realize artificial gauge fields in a transmon ladder. 
A theoretical challenge is now to extend these results to a larger
number of legs for the Hall effect, or so investigate if similar universal relations would also exist for other transport quantities such as the thermopower or Nernst effect~\cite{stafford_thermop}.  

\begin{acknowledgments}
TG would like to thank B. Altshuler for an interesting discussion. R.C. would like to thank M. Filippone for interesting insights. The authors thank N. Laflorencie for data sharing. 
This research was supported in part by the Swiss National Science Foundation under grants 200020-188687 and 200020-219400. T.G., R.C. and E.O.  would like to thank the Institut Henri Poincaré (UAR 839 CNRS-Sorbonne Universit\'e) and the LabEx CARMIN (ANR-10-LABX-59-01) for their support. TG would like to thank the program on ``New Directions in far from Equilibrium Integrability and beyond'' and the Simons Center for Geometry and Physics, Stony Brook University at which some of this work was completed, for support and hospitality.

\end{acknowledgments}


%
\end{cbunit}

\newpage 
\widetext
\begin{cbunit}
\begin{center}
\textbf{\large Supplemental Material for ''Hall response in interacting bosonic and fermionic ladders'' }
\end{center}
\setcounter{equation}{0}
\setcounter{figure}{0}
\setcounter{table}{0}
\setcounter{page}{1}
\makeatletter
\renewcommand{\theequation}{S\arabic{equation}}
\renewcommand{\thefigure}{S\arabic{figure}}
\renewcommand{\bibnumfmt}[1]{[S#1]}
\renewcommand{\citenumfont}[1]{S#1}
\onecolumngrid 
\section{Band curvature in Tomonaga-Luttinger liquids}\label{app:curvature}
\subsection{single component case}
We consider first a single component Tomonaga-Luttinger liquid \cite{giamarchi_book_1d}. The four possible  cubic order terms  in $\Pi$ and $\partial_x \phi$ are
\begin{eqnarray}\label{eq:cubic}
  H^{(3)} = \int dx \left[\alpha (\partial_x \phi)^3 + \beta \pi \Pi (\partial_x \phi)^2 + \gamma (\pi \Pi)^2 \partial_x \phi + \delta (\pi \Pi)^3\right].
\end{eqnarray}
Under particle-hole transformation, $\phi \to -\phi$ and $\Pi \to -\Pi$, so all cubic terms vanish in a particle-hole symmetric case. 

\subsubsection{Parity invariance}
Under a parity transformation $P$, we have $P\psi(x)P=\psi(-x)$, so
\begin{equation}
  e^{i k_F x} P \psi_R(x) P + e^{-i k_F x} P \psi_L(x) P = e^{-i k_F x} \psi_R(-x) + e^{i k_F x} \psi_L(-x),
\end{equation}
and $P\psi_R(x)P=\psi_L(-x)$, $P\psi_L(x) P=\psi_R(-x)$. In terms of the dual fields,
\begin{eqnarray}
  P\theta(x)P&=&\theta(-x), \\
  P \phi(x) P &=& - \phi(-x),
\end{eqnarray}
and by taking the derivatives, $P\Pi(x)P = -\Pi(-x)$ and $P \partial_x \phi(x) P =\partial_x \phi(-x)$. So if our Hamiltonian is parity invariant, we must have $\beta=\delta=0$ in Eq.~(\ref{eq:cubic}).
\subsubsection{Time reversal invariance}
In the case of time reversal $T$, since time reversal is anti unitary, we have $T^{-1} i T = -i$ and since $T^{-1} \psi(x) T = \psi(x)$
\begin{equation}
   e^{-i k_F x}  T^{-1} \psi_R(x) T + e^{i k_F x}  T^{-1} \psi_L(x) T = e^{i k_F x} \psi_R(x) + e^{-i k_F x} \psi_L(x),
 \end{equation}
 indicating that time reversal  exchanges right movers and left movers. Using the bosonized expressions, and taking into account anti unitarity, we must have
 \begin{eqnarray}
   T^{-1} \theta(x) T &=& -\theta(x),\\
   T^{-1} \phi(x) T &=& \phi(x),
 \end{eqnarray}
 so that $T^{-1} \Pi(x) T = -\Pi(x)$ and $T^{-1} \partial_x \phi T = \partial_x \phi(x)$. If the Hamiltonian has time-reversal symmetry, we also find $\beta=\delta=0$ in Eq.~(\ref{eq:cubic}).

 In conclusion, in a system without particle-hole symmetry but with 
 parity or time-reversal invariance, the most general cubic terms 
 are given by Eq.~(3) of the paper. 
 \subsection{Two component Tomonaga-Luttinger liquid}
In the case of the two component system, time reversal symmetry becomes $T^{-1}\phi_\nu T=\phi_\nu$ and $T^{-1}\Pi_\nu T=-\Pi_\nu$ for $\nu=\rho,\sigma$. If the two chains are equivalent, under exchange of the chains, $\phi_\sigma \to -\phi_\sigma$ and $\Pi_\sigma \to -\Pi_\sigma$ is another symmetry.
The most general cubic term to add to the bosonized
Hamiltonian is then of the form
\begin{eqnarray}
H_{\mathrm{curvature}}= \sum_{\sum n_j=3} \int dx \alpha_{n_1 n_2 n_3 n_4} (\pi \Pi_\rho)^{n_1} (\pi \Pi_\sigma)^{n_2} \nonumber \\ \times (\partial_x \phi_\rho)^{n_3} (\partial_x \phi_\sigma)^{n_4},
\end{eqnarray}
with the constraints $n_1+n_2$ even from time reversal invariance, and $n_2+n_4$ even from symmetry under chain exchange.
With the help of the symmetries, the cubic term reduces to
\begin{eqnarray}\label{eq:2comp-cubic}
H_{\mathrm{curvature}}&=& \int dx \left[ a_1 (\pi \Pi_\rho)^2 \partial_x \phi_\rho + a_2 (\partial_x \phi_\rho)^3 + a_3 (\pi \Pi_\sigma)^2 \partial_x \phi_\rho \right. \nonumber \\ &&\left.+ a_4 (\partial_x \phi_\rho)^2 \partial_x \phi_\sigma + a_5 \pi^2 \Pi_\rho \Pi_\sigma \partial_x \phi_\sigma \right],
\end{eqnarray}
where we have set $a_1=\alpha_{2010}$, $a_2=\alpha_{0030}$, $a_3=\alpha_{0210}$, $a_4=\alpha_{0012}$, $a_5=\alpha_{1101}$ to lighten the notation. 
Eq.~(9) of the letter (with $A_\rho=A_\sigma=0$) is of the the form (\ref{eq:2comp-cubic}) with 
\begin{eqnarray}
&& a_1=\gamma/\sqrt{2} ;\quad a_2=\alpha/\sqrt{2} \\ 
&& a_3=\gamma/\sqrt{2}; \quad a_4=3\alpha/\sqrt{2} \\ 
&& a_5 = \sqrt{2}\gamma 
\end{eqnarray}

\subsection{Microscopic expressions of the band curvature coefficients}
The coefficients $a_{1,2,3,4}$ can be related with $u_{s,a}$ and $K_{s,a}$ by considering the effect of a change of average total density $\rho_t \to \rho_t +\Delta\rho$ equivalent to $\partial_x \phi_\rho \to \partial_x \phi_\rho -\pi \sqrt{2} \Delta\rho/\pi$ on the Hamiltonian \cite{matveev_equilibration_2013,pereira_dynamical_2007,pereira_spin-charge_2010}. Under such change, $H_{\mathrm{curvature}}$ adds a contribution
\begin{eqnarray}
   \frac{-\pi \Delta \rho}{2} \int dx \left[ a_1 (\pi \Pi_\rho)^2 + 3 a_2 (\partial_x \phi_\rho)^2 + a_3 (\pi \Pi_\sigma)^2 + a_4 (\partial_x \phi_\sigma)^2  \right]
\end{eqnarray}
to the quadratic Hamiltonian. Since the resulting Hamiltonian must coincide with the bosonized Hamiltonian of a system of density $\rho_0+\Delta\rho$, we must have the relations
\begin{eqnarray}
a_1 &=& -\frac{\partial}{\partial \rho_t} \left(\frac{u_\rho K_\rho}{\pi^2 \sqrt{2}} \right) \\
a_2 &=& -\frac{\partial}{\partial \rho_t} \left(\frac{u_\rho }{3 \pi^2 K_\rho \sqrt{2}}  \right) \\
a_3 &=& -\frac{\partial}{\partial \rho_t} \left(\frac{u_\sigma K_\sigma}{\pi^2 \sqrt{2}} \right) \\
a_4 &=& -\frac{\partial}{\partial \rho_t} \left(\frac{u_\sigma}{3 \pi^2 K_\sigma \sqrt{2}}  \right)
\end{eqnarray}
Noting that $\rho_t=2\rho_0$, it is straightforward to verify that for $u_\rho=u_\sigma=u$ and $K_\rho=K_\sigma=K$,  $a_{1,2,3,4}$ reduce to the coefficient in Eq.~(9) of the letter.
In the case of a Galilean invariant system, a further relation \cite{pereira_spin-charge_2010} can be obtained by considering the total particle current $j_s=\frac{\sqrt{2}}{\pi} \partial_x \phi_\rho$.
By commuting $\phi_\rho$ with the full Hamiltonian, the particle current is obtained as
\begin{eqnarray}
j_s (x) =\sqrt{2} \left[u_\rho K_\rho \Pi_s + \pi (2 a_1 \Pi_s \partial_x \phi_\rho + a_5 \Pi_\sigma \partial_x \phi_\sigma)\right],
\end{eqnarray}
and in a Galilean invariant system, the current
\begin{equation}
J_s=\int dx j_s (x),
\end{equation}
is proportional to the momentum $P$. Since the momentum operator $P$ is the infinitesimal generator of translations \cite{itzykson_zuber}, given the canonical commutation relation, it is
\begin{equation}
    P=\int dx \left[-\frac{\pi \rho_t}{\sqrt{2}} \Pi_s + \Pi_s \partial_x \phi_\rho + \Pi_\sigma \partial_x \phi_\sigma \right].
\end{equation}
As a result, in a Galilean invariant system, we must have $a_5=2a_1$ and $u_\rho K_\rho \propto \rho_t$. In turn, this implies that $a_1$ and $a_5$ are unrenormalized by interactions and independent of density. Since the Hall resistivity $R_H$ is proportional to $a_5/(u_\rho K_\rho)$, we conclude that in the Galilean invariant case, the Hall resistance is proportional to the inverse of the density, and independent of interaction.
In the absence of Galilean invariance, we can relate $a_5$ to the variation of the ground state energy under application of Aharonov-Bohm fluxes $\Phi_1$ in chain $1$ and $\Phi_2$ in chain 2 in the presence of a density imbalance $\Delta \rho_a$. Under such perturbations, we have
\begin{eqnarray}
\langle \pi \Pi_s \rangle &=& \frac{\Phi_1+\Phi_2}{\sqrt{2} L}, \\
\langle \pi \Pi_\sigma \rangle &=& \frac{\Phi_1-\Phi_2}{\sqrt{2} L}, \\
\langle \partial_x \phi_\sigma \rangle &=& -\frac{\pi \Delta\rho_a}{\sqrt{2}},
\end{eqnarray}
yielding a shift in ground state energy
\begin{eqnarray}
\Delta E = -a_5\frac{\pi \Delta \rho_a (\Phi_1^2-\Phi_2^2)}{2 \sqrt{2} L},
\end{eqnarray}
that can be matched with the Taylor expansion to third order of the ground state energy to obtain
\begin{eqnarray}\label{eq:a5-nonpert}
a_5=-\frac{\sqrt{2} L}{\pi} \frac{\partial^3 E}{\partial^2 \Phi_1 \partial (\Delta \rho_a)} = \frac{\sqrt{2} L}{\pi} \frac{\partial^3 E}{\partial^2 \Phi_2 \partial (\Delta \rho_a)},
\end{eqnarray}
or in terms of particle densities $\rho_1$ in chain 1, and $\rho_2$ in chain 2,
\begin{eqnarray}
    a_5=-\frac{L}{\pi \sqrt{2}} \left( \frac{\partial^3 E}{\partial^2 \Phi_1 \partial \rho_1} - \frac{\partial^3 E}{\partial^2 \Phi_1 \partial \rho_2}  \right).
\end{eqnarray}
When the chains are decoupled, $\partial^3E/(\partial_2\varphi_1\partial\rho_2)=0$, and  this non-perturbative expression of $a_5$ reduces to the derivative of charge stiffness with respect to density that we obtained using perturbation theory. A formula analogous to Eq.~(\ref{eq:a5-nonpert}), relating the Hall resistivity to a third derivative of ground state energy,  has been derived in  \cite{prelovsek_hall_1999}. In our notations,
it reads
\begin{equation}
R_H \propto \frac{\frac{\partial^3 E}{\partial A_s \partial A_a \partial V_H}}{\frac{\partial^2 E}{\partial^2 V_H}}
\end{equation}
Now, if we consider a small variation of $V_H$, the variation of the imbalance will be $\Delta \rho_a = (\partial^2 E/\partial V_H^2)  \Delta V_H$, injecting the relation in the shift of the ground state energy gives back the relation of  \cite{prelovsek_hall_1999}.

\section{band curvature terms and gauge invariance}\label{app:gauge}
In the presence of band curvature terms \cite{haldane_bosonisation}, the Hamiltonian of a Tomonaga-Luttinger liquid is
\begin{eqnarray}
  \label{eq:bosonized}
  H&=&\int \frac{dx}{2\pi} \left[u K( \pi \Pi -e A)^2 + \frac u K (\partial_x \phi)^2\right] \nonumber \\&&+ \int dx \left[\alpha (\partial_x \phi)^3 +   \gamma (\pi \Pi-e A)^2 \partial_x \phi\right].
\end{eqnarray}
Indeed, in the absence of the vector potential, we recover the usual expressions of the Tomonaga-Luttinger Hamiltonian with band curvature \cite{haldane_bosonisation,matveev_equilibration_2013}. Under a gauge transformation, $A \to A+ \nabla \chi$, while the fermion annihilation operators
\begin{eqnarray}\label{eq:bosonized-fermions}
  \psi_{R} &=& \frac{e^{i(\theta -\phi)}}{\sqrt{2\pi a_0}} \eta, \\
   \psi_{L} &=& \frac{e^{i(\theta +\phi)}}{\sqrt{2\pi a_0}} \eta,
\end{eqnarray}
transform \cite{messiah_mecanique_quantique} as $\psi_{\nu} \to e^{i e \chi} \psi_{\nu}$, yielding $\theta \to \theta+ e\chi$, with $\phi$ invariant. This implies that  the combination $\pi \Pi -e A$ satisfies gauge invariance, ensuring that Eq.~(\ref{eq:bosonized}) is gauge invariant.

\section{Perturbative calculation of the Hall imbalance in the bosonic ladder}
\label{sec:bosonized-ladder}

The denominator of Eq.~(9) is $1+O(\gamma,\alpha)$. Since the numerator is $O(\gamma,\alpha)$ its contribution to $\langle P_H \rangle$ is thus $O(\gamma^2,\alpha\gamma,\alpha^2)$ and can be neglected. 
Expanding the exponential of the numerator to first order in $H_{curv.}$ in Eq.~(9) of the letter gives

\begin{equation}\label{eq:hallaverage-s}
\begin{split}
  \langle P_H^{(1)}\rangle &= \frac{\sqrt{2}}\pi \int_0^\beta d\tau \int_0^L dx   \left\langle T_\tau  H_{curv.}(\tau) \partial_x \phi_\sigma(x,0) \right\rangle_{H_\rho+H_\sigma+H_{hop.}} \\
                     &=  \frac {2\gamma} {\pi} \int_0^L dy   \int_0^\beta d\tau \int_0^L dx     \langle (\pi \Pi_\rho - e A_\rho) \rangle_{H_\rho} \left\langle T_\tau \left[   (\pi \Pi_\sigma - e A_\sigma)  \partial_y \phi_\sigma       \right](y,\tau)\partial_x \phi_\sigma(x,0) \right\rangle_{H_\sigma+H_{hop.}},
\end{split}
\end{equation}

where in the second line of Eq.~(\ref{eq:hallaverage-s}), we have taken into account the particle-hole symmetry of $H_\rho+H_\sigma+H_{hop.}$. As a result, the term of order zero and all contributions cubic or linear  in $\partial_x\phi_\rho$ in the first order term vanish, leaving only the contribution of $(\pi\Pi_\rho -A A_\rho)(\pi \Pi_\sigma -e A_\sigma) \partial_x \phi_\sigma$ to consider. Moreover, since $H_\rho$ and $H_\sigma+H_{hop.}$ decouple, we can average separately $\pi\Pi_\rho -e A_\rho$ and $(\pi\Pi_\sigma - e A_\sigma)\partial_x \phi_\sigma$. 
Using the definition of the current, 
\begin{equation}
 \frac{\langle P_H^{(1)}\rangle}{L} =\frac{\gamma \langle j_\rho \rangle \sqrt{2}}{e u_\rho K_\rho}\int_0^\beta d\tau \int_0^L dx \left\langle T_\tau \left[   (\pi \Pi_\sigma - e A_\sigma)  \partial_x \phi_\sigma       \right](x,\tau)\partial_x \phi_\sigma(0,0) \right\rangle_{H_\sigma+H_{hop.}},
\end{equation}
\paragraph{Meissner phase}
In the limit of small flux, in the Meissner phase, we can separate the expectation value into 
$\langle T_\tau \Pi_\sigma(x,\tau)  \partial_x \phi_\sigma (x,\tau) \partial_x\phi_\sigma(0,0) \rangle_{A_\sigma=0}$ and $A_\sigma \langle T_\tau \partial_x \phi_\sigma (x,\tau) \partial_x\phi_\sigma(0,0) \rangle_{A_\sigma=0}$. The first term vanishes by the particle-hole symmetry of the sine-Gordon Hamiltonian ($\Pi_\sigma \to -\Pi_\sigma,\phi_\sigma \to -\phi_\sigma$), as long as our system remains in the commensurate phase. That leaves us only with the second term
\begin{equation}
 \frac{\langle P_H^{(1)}\rangle}{L} =\frac{\gamma \langle j_\rho \rangle \sqrt{2} A_\sigma }{ u_\rho K_\rho} \int_0^\beta d\tau \int_0^L dx \left\langle T_\tau \partial_x \phi_\sigma(x,\tau) \partial_x \phi_\sigma(0,0) \right\rangle_{H_\sigma+H_{hop.}},
\end{equation}
which is proportional to the static susceptibility
\begin{equation}
\chi_{\sigma\sigma} (q=0,\omega_n=0) = -\frac 2 {\pi^2} \int_0^\beta d\tau \int_0^L dx \left\langle T_\tau \partial_x \phi_\sigma(x,\tau) \partial_x \phi_\sigma(0,0) \right\rangle_{H_\sigma+H_{hop.}}, 
\end{equation}
so that 
\begin{equation}
\frac{\langle P_H^{(1)}\rangle}{L} =\frac{\pi^2 \gamma \langle j_\rho \rangle B a }{2 u_\rho K_\rho} \langle j_\rho \rangle \chi_{\sigma\sigma}    
\end{equation}

An exact expression of the static susceptibility $\chi_{\sigma\sigma}$ is obtained by adding a perturbation $\lambda \sqrt{2} \partial_x \phi$ to $H_\sigma+H_{hop.}$, completing the square and shifting $\partial_x \phi$ by a constant. One finds $\chi_{\sigma\sigma}=\frac{2K_\sigma}{\pi u_\sigma}$, and in the Meissner phase
\begin{equation}
\frac{\langle P_H^{(1)}\rangle}{L} =\frac{\pi \gamma K_\sigma a }{u_\sigma u_\rho K_\rho} \langle j_\rho \rangle B 
\end{equation}
Numerical simulation \cite{buser_hall_ladder} show a dependence of Hall polarization on flux, density and filling. 
\paragraph{Vortex phase}
When we are in the vortex phase, $\langle \Pi_\sigma \rangle \ne 0$, and as a result, the evaluation of $\langle T_\tau \Pi_\sigma(x,\tau)  \partial_x \phi_\sigma (x,\tau) \partial_x\phi_\sigma(0,0) \rangle$ is slightly different. Since the fixed point Hamiltonian in the Meissner phase is 
\begin{equation}
H_\sigma^* = \int \frac{dx}{2\pi} \left[ u_\sigma^* K_\sigma^* (\pi \Pi_\sigma -\pi \langle \Pi_\sigma\rangle)^2 + \frac{u^*_\sigma}{K^*_\sigma} (\partial_x\phi_\sigma)^2 \right], 
\end{equation}
the symmetry becomes $\Pi_\sigma \to 2\langle \Pi_\sigma \rangle -\Pi_\sigma$ and $\phi_\sigma \to -\phi_\sigma$, and 
 \begin{equation}
  \langle T_\tau \Pi_\sigma(x,\tau)  \partial_x \phi_\sigma (x,\tau) \partial_x\phi_\sigma(0,0) \rangle = \langle \Pi_\sigma \rangle \langle T_\tau \partial_x \phi_\sigma (x,\tau) \partial_x\phi_\sigma(0,0) \rangle. 
 \end{equation}
We find 
\begin{equation}
\begin{split}
 \frac{\langle P_H^{(1)}\rangle}{L} =\frac{\gamma \langle j_\rho \rangle \sqrt{2}}{e u_\rho K_\rho} (\pi \langle \Pi_\sigma \rangle - e A_\sigma) \int_0^\beta d\tau \int_0^L dx \left\langle T_\tau \partial_x \phi_\sigma (x,\tau)\partial_x \phi_\sigma(0,0) \right\rangle_{H_\sigma+H_{hop.}},
 &= \frac{\gamma \langle j_\rho \rangle \sqrt{2}}{e u_\rho K_\rho} (\pi \langle \Pi_\sigma \rangle - e A_\sigma) \frac{\pi K^*_\sigma}{u^*_\sigma}.
 \end{split}
\end{equation}
For large flux, $(\pi \langle \Pi_\sigma \rangle - e A_\sigma) \to 0$, and the Hall imbalance vanishes. In that limit, the chains are decoupled by the applied flux. 
The Meissner and the Vortex phase are separated by a commensurate-incommensurate transition \cite{orignac01_meissner}. 
The vicinity of the commensurate-incommensurate transition requires a specific treatment, for instance by considering the Luther-Emery limit $K_\sigma=1/2$. 
\subsection{Hall voltage}
To compute the  the Hall Voltage we add  a potential difference $V_H$ between the two chains  \cite{prelovsek_hall_1999},
\begin{equation}
  -\frac{eV_H}{\pi\sqrt{2}} \int dx \partial_x \phi_\sigma,
\end{equation}
and we search for $V_H$ such that $\langle \partial_x\phi_\sigma\rangle =0 $ in first order perturbation theory. The condition becomes
\begin{eqnarray}
\int d\tau dx \left\langle T_\tau \left[\sqrt{2}\gamma (\pi\Pi_\rho - e A_\rho)(x,\tau) (\pi\Pi_\sigma - e A_\sigma)(x,\tau) \partial_x\phi_\sigma(x,\tau) -\frac{eV_H}{\pi\sqrt{2}} \partial_x\phi_\sigma(x,\tau)  \right] \partial_x\phi_\sigma(0,0) \right\rangle = 0  
\end{eqnarray}
In the Meissner phase, the Hall voltage is
\begin{equation}\label{eq:hall-voltage-bosons-s}
  V_H=-\frac{\pi^2 \gamma \langle j_s \rangle}{e u_\rho K_\rho} B a = 
  \langle j_s \rangle \frac{B a}{2e} \frac 1 {u_\rho K_\rho} \frac{\partial}{\partial \rho_0} (u_\rho K_\rho),
\end{equation}
where we have used the relation \cite{matveev_equilibration_2013}
\begin{equation}
\gamma=-\frac{\partial}{\partial\rho_0} \left(\frac{u_\rho K_\rho}{2\pi^2} \right)
\end{equation}
Since the longitudinal current density $I_x=\langle j_s\rangle/2$ this formula reduces to
\begin{equation}\label{eq:hall-resistance-bosons-s}
R_H=\frac{a}{e u_\rho K_\rho} \frac{\partial}{\partial \rho_0} (u_\rho K_\rho).
\end{equation}
In a system with Galilean invariance, the stiffness $u_\rho K_\rho = \frac{\pi \rho_0}{m}$ is unrenormalized \cite{haldane_bosons}, and Eq.~(\ref{eq:hall-resistance-bosons-s}) simplifies to 
\begin{equation}\label{eq:rh_naive-s}
R_H=\frac{a}{e\rho_0}, 
\end{equation}
which is the usual formula for the Hall resistance \cite{ziman_solid_book} in a metal. 
Numerical simulations at low flux, show (Figs. 5a and 5b of  \cite{buser_hall_ladder}) a Hall voltage multiplied by density behaving linearly with the flux, in agreement with Eq.~(\ref{eq:rh_naive-s}).
For realization of ladders coming from synthetic dimensions, an interchain interaction exists in addition to the tunnelling. For incommensurate fillings, such an interaction comprises forward scattering $\partial_x \phi_1 \partial_x \phi_2$ that renormalizes  $u_\rho/K_\rho$ and $u_\sigma/K_\sigma$ but not $u_\rho K_\rho$ or $u_\sigma K_\sigma$, and backward scattering $\cos 2(\phi_1-\phi_2) $ that renormalizes $K_\sigma$ but not $K_\rho$.
Thus, at the perturbative level, when backward scattering is irrelevant, $u_\rho K_\rho$ is not affected by interactions and Eq.~(\ref{eq:hall-resistance-bosons-s}) remains applicable even in the presence of interchain interaction as long as the system is in the Meissner phase. Furthermore, if the system has Galilean invariance, the Hall resistance will be given by Eq.~(\ref{eq:rh_naive-s}).  When interchain backward scattering becomes relevant, the system enters a charge density wave phase \cite{orignac_2chain_bosonic,Orignac2017,Citro2018} in which $\chi_{\sigma\sigma}=0$. 
In such phase, $\langle P_H \rangle^{(1)}=0$ and no Hall effect is present.  
An important case in which Galilean invariance is absent is when a periodic lattice is present, allowing the formation of a Mott insulator \cite{giamarchi_mott_review}. 
In the vicinity of commensurate filling, the velocity $u_\rho \sim (\rho_0-\rho_{0,c})$ while $K_\rho$ is renormalized to a constant value. This 
implies $R_H$ diverges as $1/(\rho_0- \rho_{0,c})$. Since the divergence of the Hall resistance results from a vanishing $u_\rho$ it is a signature of the breakdown of Tomonaga-Luttinger theory in the vicinity of $\rho_{0,c}$. To determine the actual behavior of $R_H$ in the immediate vicinity of $\rho_{0,c}$ demands a full study of the commensurate-incommensurate quantum critical point \cite{japaridze_cic_transition,pokrovsky_talapov_prl,sachdev_2d_qcp_spins}. Both the Mott transition and the band-filling transition have been studied in the case of a ladder of hard core bosons \cite{Crepin2011}, making such systems a potential testbed for the divergence of the Hall resistance. Simultaneous zeros or divergences of the TL exponent $K_\rho$ and the velocity $u_\rho$ are also possible \cite{cabra_instabilityLL}, and would give rise to a different prefactor in the divergence of $R_H$.

\section{Perturbative calculation of Hall imbalance in the fermionic ladder}
\label{sec:fermions}
The calculation follows the same outline as in the fermionic case, but there are a few significant differences due to differences in the low-energy 
theory describing the antisymmetric modes. 
\subsection{Low energy Hamiltonian}
The full second quantized Hamiltonian reads
\begin{multline}
  \label{eq:hamiltonian-s}
  H=\sum_{p=1,2} \int dx \frac{\psi^\dagger_p (x)}{2m}\left(\frac \hbar i \nabla - e A_p\right)^2 \psi_p(x)-\mu \psi^\dagger_p(x) \psi_p(x)
-t_\perp \int dx (\psi^\dagger_1 \psi_2 + \psi^\dagger_2 \psi_1) \\ 
     + U \int dx \rho_1(x) \rho_2(x) 
     + \frac 1 2 \int dx V(x-x') [\rho_1(x)\rho_1(x')+\rho_2(x)\rho_2(x')]
     \end{multline}
where $\psi_p$ annihilates a fermion in the chain $p$, $\rho_p =\psi^\dagger_p \psi_p$,  $A_p$ is the vector potential in the Landau gauge along the chain $p$, $\mu$ the chemical potential, $m$ the mass, $e$ the fermion charge, $V(x)$  the short ranged-intrachain interaction, $U$ the interchain interaction and $t_\perp$ the interchain hopping. With $V(x)=0$, Eq.~(\ref{eq:hamiltonian}) reduces to the Gaudin-Yang \cite{gaudin_fermions,yang_fermions} model in magnetic field. 

Retaining only momenta around the Fermi points, we rewrite the Hamiltonian~(\ref{eq:hamiltonian}) in the form
\begin{eqnarray}
  \label{eq:linearized-ham}
  H&=&H_R+H_L+H_{tunn.} + H_{int.} \\
  H_R&=&\sum_{p=1,2} \int dx v_F \psi^\dagger_{R,p} (-i \partial_x - e A_p) \psi_{R,p}  + \frac 1 {2m} \int dx  \psi^\dagger_{R,p} (i \partial_x + e A_p)^2\psi_{R,p}, \\
  H_L&=&\sum_{p=1,2} \int dx v_F \psi^\dagger_{L,p} (i \partial_x  + e A_p) \psi_{L,p} + \frac 1 {2m} \int dx  \psi^\dagger_{L,p} (i \partial_x + e A_p)^2\psi_{L,p}, \\
  H_{tunn.} &=& -t_\perp \sum_{\nu=R,L} \int dx (\psi^\dagger_{\nu,1} \psi_{\nu,2} + \psi^\dagger_{\nu,2} \psi_{\nu,1}), \\
  H_{int} &=& U \int dx [(\rho_{R,1}+\rho_{L,1}) (\rho_{R,2}+\rho_{L,2})]  + U \int dx [\psi^\dagger_{R,1} \psi^\dagger_{L,2} \psi_{R,2} \psi_{L,1} +\psi^\dagger_{R,2} \psi^\dagger_{L,1} \psi_{R,1} \psi_{L,2} ] \nonumber\\
             && + [\hat{V}(0)-\hat{V(2k_F)}] \int dx (\rho_{R,1} \rho_{L,1} + \rho_{R,2} \rho_{L,2}) +\frac{\hat{V}(0)} 2 \int dx (\rho_{R,1}^2 +\rho_{R,2}^2 +\rho_{L,1}^2 +\rho_{L,2}^2),
\end{eqnarray}
with $\psi_p(x)=e^{i k_F x} \psi_{Rp}(x)+e^{-i k_F x} \psi_{Lp}(x)$, with $k_F=\pi \rho_0$ the Fermi wavevector and $v_F=k_F/m$ the Fermi velocity.
We have defined
\begin{equation}
\hat{V}(k)=\int dx V(x) e^{-ikx},
\end{equation}
the Fourier transform of the intrachain interaction.

\subsection{Change of basis}
It is convenient to rewrite (\ref{eq:linearized-ham}) in a bonding-antibonding band basis \cite{nersesyan_2ch}
\begin{eqnarray}
  \label{eq:bonding}
  \psi_{\nu,p}=\frac{\psi_{\nu,0}+(-1)^{p-1}\psi_{\nu,\pi}}{\sqrt{2}},
\end{eqnarray}
that diagonalizes $H_{tunn.}$.

Following a standard procedure, one can introduce $A_{\rho,\sigma}=(A_1\pm A_2)/2$
and bosonize the Hamiltonian \cite{nersesyan_2ch,giamarchi_book_1d} expressing it in terms of the charge and spin basis $(\rho,\sigma)$,

In the bonding-antibonding basis introduced in Eq.~(\ref{eq:ham-bonding}) with the vector potentials $A_{\rho,\sigma}=\frac{A_1\pm A_2}{2}$, the non-interacting Hamiltonian takes the form
\begin{equation}
\begin{split}
  \label{eq:ham-bonding}
  H_R+H_L+H_{tunn.}
  &= v_F  \sum_{j=0,\pi} \int dx  \left[\psi^\dagger_{R,j} (-i \partial_x -e A_\rho) \psi_{R,j} +\psi^\dagger_{L,j} (i \partial_x +e A_\rho) \psi_{L,j} \right] \\
    &+\frac 1 {2m}  \sum_{j=0,\pi \atop \nu=R,L } \int dx \psi^\dagger_{\nu,j} [(i \partial_x + e A_\rho)^2 + (e A_\sigma)^2] \psi_{\nu,j} 
    - e v_F A_\sigma \int dx (\psi^\dagger_{R,0} \psi_{R,\pi}   -\psi^\dagger_{L,0} \psi_{L,\pi}  +\hc ) \\
  &+\frac{eA_\sigma} m \sum_{\nu=R,L} \int dx \left[\psi^\dagger_{\nu,0} (i\partial_x + e A_\rho)  \psi_{\nu,\pi}   + \psi^\dagger_{\nu,\pi} (i\partial_x + e A_\rho) \right].
  \end{split}
\end{equation}
We introduce a bosonized representation \cite{giamarchi_book_1d},
\begin{equation}
  \label{eq:bosonized-bonding}
  \psi_{R,j} =\frac{e^{i(\theta_j -\phi_j)}}{\sqrt{2\pi a_0}} \quad \psi_{L,j} =\frac{e^{i(\theta_j +\phi_j)}}{\sqrt{2\pi a_0}}
\end{equation}
where
\begin{eqnarray}
 \pi \Pi_j=\partial_x \theta_j, \\
  \label{eq:commutator}
  [\phi_j(x),\Pi_{j'}(y)]=i \delta_{jj'}\delta(x-y),
\end{eqnarray}
with $j,j'\in \{0,\pi\}$, and $a_0$ a short distance cutoff. It is convenient to use new boson fields \cite{nersesyan_2ch,orignac97_ladder}
\begin{eqnarray}
  \label{eq:rotation}
  \phi_{\rho}=\frac{\phi_0+\phi_\pi}{\sqrt{2}} \;  \theta_{\rho}=\frac{\theta_0+\theta_\pi}{\sqrt{2}} , \\
  \phi_{\sigma}=\frac{\phi_0-\phi_\pi}{\sqrt{2}} \;  \theta_{\sigma}=\frac{\theta_0-\theta_\pi}{\sqrt{2}} ,
\end{eqnarray}
In terms of the new fields, we have
\begin{eqnarray}
P_H&=&\sum_{\nu=R,L} \int dx (\psi^\dagger_{\nu,0} \psi_{\nu,\pi} + \psi^\dagger_{\nu,\pi} \psi_{\nu,0} )
    =\int dx \frac{2\cos \sqrt{2} \theta_\sigma \cos \sqrt{2} \phi_\sigma}{\pi a_0},
\end{eqnarray}

In the bosonized language the non-interacting Hamiltonian is rewritten as
\begin{align}\label{eq:nonint-bosonized}
\begin{split}
  H_R + H_L
  &=\frac{v_F}{2\pi}\int dx  \left[(\pi \Pi_\sigma)^2 +(\partial_x \phi_\sigma)^2 + (\pi \Pi_\rho - e\sqrt{2} A_\rho)^2 + (\partial_x \phi_\rho)^2\right]  \\
  &- \frac{2 ev_F A_\sigma} {\pi a_0} \int dx \sin \sqrt{2} \theta_\sigma \sin \sqrt{2} \phi_\sigma  -\int dx \frac{(e A_\sigma)^2}{\pi m\sqrt{2}} \partial_x \phi_\rho+ \frac{\sqrt{2} t_\perp}\pi \int dx \partial_x \phi_\sigma \\ 
     &-\frac 1 {6\pi m\sqrt{2}} \int dx \left[ 3 (\pi \Pi_\rho -e \sqrt{2} A_\rho)^2\partial_x \phi_\rho + (\partial_x \phi_\rho)^3+3((\pi \Pi_\sigma)^2 +(\partial_x\phi_\sigma)^2) \partial_x \phi_\rho + 6 (\pi \Pi_\rho -e \sqrt{2} A_\rho) \pi \Pi_\sigma \partial_x \phi_\sigma \right]\\
      &+\frac{\sqrt{2} e A_\sigma}{\pi m a_0} \int dx \left[\sin \sqrt{2}\theta_\sigma \sin \sqrt{2}\phi_\sigma \partial_x\phi_\rho - \cos \sqrt{2} \theta_\sigma \cos \sqrt{2} \phi_\sigma (\pi \Pi_\rho -e \sqrt{2} A_\rho) \right],
     \end{split}
\end{align}
while interactions take the form
\begin{align}
\begin{split}
  H_{int.}
  &=\int dx \left[\frac{2U + 2\hat{V}(0)-\hat{V}(2k_F)}{4\pi^2} (\partial_x \phi_\rho)^2 +\frac{\hat{V}(2k_F)}{4\pi^2} (\Pi_\rho -e\sqrt{2}A_\rho)^2 + \frac{\hat{V}(0)}{4\pi^2} (\pi \Pi_\sigma)^2 +\frac{\hat{V}(0)-2 U}{4\pi^2} (\partial_x \phi_\sigma)^2 \right] \\
  &+\frac{V(0)-V(2k_F)}{(2\pi a_0)^2} \int dx \cos \sqrt{8} \theta_\sigma + \frac{2U +\hat{V}(0)-\hat{V}(2k_F)}{(2\pi a_0)^2} \int dx \cos \sqrt{8} \phi_\sigma,
  \end{split}
\end{align}
and the Hall polarization is
\begin{eqnarray}
  \label{eq:imbalance-bosonized-s}
  P_H=2\int \frac{dx}{\pi a_0} \cos \sqrt{2} \theta_\sigma \cos \sqrt{2} \phi_\sigma.
\end{eqnarray}
The ground state of the Hamiltonian $H_0+H_{int.}$ for $m\to +\infty$ and $A_\rho=A_\sigma=0$ \cite{nersesyan_2ch,orignac97_ladder} is well known. The field $\theta_\sigma$ is long range order, with expectation value $\langle \theta_\sigma \rangle =\pi/\sqrt{8}$, while the field $\theta_\rho$ remains gapless. Upon application of $A_\rho \ne 0$, a diamagnetic current $\langle j_s\rangle =\sqrt{2} u_\rho K_\rho  (\pi \langle \Pi_\rho\rangle -e \sqrt{2} A_\rho)/\pi$ proportional to $A_\rho$ is generated, while in a magnetic field $B$,  $A_\sigma = Ba/2$. Eq.~(\ref{eq:nonint-bosonized}), contains a term proportional to $ B j_s$ in the last line.
To first order in $1/m$ it gives a contribution to the Hall polarization
\begin{eqnarray}
  \langle P_H \rangle &=& \frac{\pi e^2 B a \langle j_s \rangle}{u_\rho K_\rho} \int \frac{dx d\tau}{(\pi a_0)^3} \langle T_\tau \cos \sqrt{2} \theta_\sigma(x,\tau)  \cos \sqrt{2} \phi_\sigma(x,\tau)   \cos \sqrt{2} \theta_\sigma(0,0)  \cos \sqrt{2} \phi_\sigma(0,0) \rangle,
\end{eqnarray}
where the correlator is calculated in the unperturbed ($m=+\infty, A_\rho=A_\sigma=0$) ground state. Since \cite{giamarchi_spin_flop,nersesyan_2ch} $\phi_\sigma = \tilde{\phi}_\sigma -\sqrt{2} t_\perp x/v_F$ with the Hamiltonian
\begin{eqnarray}\label{eq:spinflop}
  \tilde{H} = \int dx \left\lbrack  \frac{v_F}{2\pi} [(\pi \Pi_\sigma)^2+(\partial_x \tilde{\phi}_\sigma)^2] +\frac{V(0)-V(2k_F)}{(2\pi a_0)^2} \cos \sqrt{8} \theta_\sigma \right\rbrack,
\end{eqnarray}
we can rewrite
\begin{eqnarray}\label{eq:hall-pol-fermi}
  \langle P_H \rangle =  \frac{\pi e^2 B a \langle j_s \rangle}{u_\rho K_\rho}  \int \frac{dx d\tau}{(\pi a_0)^2} \left\langle T_\tau \cos \sqrt{2} \theta_\sigma(x,\tau)  \cos \left(\sqrt{2} [\tilde{\phi}_\sigma(x,\tau)- \tilde{\phi}_\sigma(0,0)] -\frac{2t_\perp x}{v_F} \right) \cos \sqrt{2} \theta_\sigma(0,0) \right\rangle,
\end{eqnarray}
where we have used the $U(1)$ symmetry of the Hamiltonian~(\ref{eq:spinflop}) under $\tilde{\phi}_\sigma \to \tilde{\phi_\sigma}+\lambda$.
Using a duality $\theta_\sigma \leftrightarrow \tilde{\phi}_\sigma$, the integral in Eq.~(\ref{eq:hall-pol-fermi}) is brought to the form
\begin{eqnarray}
  \int \frac{dx d\tau}{(\pi a_0)^2} \left\langle T_\tau \cos \sqrt{2} \tilde{\phi}_\sigma(x,\tau)  \cos \sqrt{2} \tilde{\phi}_\sigma(0,0)   \cos \left(\sqrt{2} [\theta_\sigma(x,\tau)- \theta_\sigma(0,0)] -\frac{2t_\perp x}{v_F} \right)\right\rangle,
\end{eqnarray}
while the sine-Gordon Hamiltonian~(\ref{eq:spinflop}) is reduced the the bosonized Hamiltonian of an easy-axis XXZ spin-1/2 chain \cite{giamarchi_book_1d}. Since the magnetization density in the hard plane of such XXZ chain is \cite{giamarchi_book_1d}
\begin{equation}
S^x(x)+iS^y(x) =\frac{e^{i\sqrt{2}\theta_\sigma}}{\pi a_0} \cos \sqrt{2} \phi_\sigma,
\end{equation}
the integral in Eq.~(\ref{eq:hall-pol-fermi})  gives the static magnetic susceptibility in the hard plane $\chi_{xx}$ of an XXZ spin-1/2 chain in the Ising phase at a wavevector $2 \tilde{t}_\perp/v_F$. \cite{caux_two-spinon_2008}
So finally,
\begin{equation}
\langle P_H \rangle = \frac{e^2 B a \langle j_s \rangle}{m u_\rho K_\rho} \chi_{xx}\left(q=\frac{2t_\perp}{v_F},\omega_n=0\right),
\end{equation}
which is formally similar to the bosonic result. In fact, since the easy axis XXZ chain is in the Ising phase but close to the isotropic point, its gap shows a Kosterlitz-Thouless essential singularity \cite{shankar_spinless_conductivite}, so with a reasonably large $t_\perp/v_F$, the susceptibility is the same as the one of the isotropic chain. Up to logarithmic corrections, the susceptibility goes as $1/v_F$, and the perturbative behavior of the imbalance in the fermionic ladder is the same as in the bosonic ladder.
In the absence of intrachain interaction, $V(x)=0$, the interaction $H_{int.}$ has a $\mathrm{SU(2)}$ symmetry, and the antisymmetric field $\theta_\sigma$ is also gapless. The expression of the Hall polarization is the same, but the magnetic susceptibility $\chi_{xx}$ is the one of the Hubbard model in a Zeeman field \cite{schulz_su2}. It still behaves as $1/v_F$ up to logarithmic corrections, so even in that high symmetry case, the behavior is the same as in the bosonic case. In our approach, interchain hopping is treated within bosonization, which is valid provided the interchain hopping is small compared with the bandwidth in an isolated chain. For interchain hopping of the order of the bandwith, a different approach in which the full noninteracting band structure is obtained first, and interactions are added perturbatively is called for. \cite{Narozhny2005}
If we now turn to the Hall resistance, we can follow the same steps as in Sec.~\ref{sec:bosonized-ladder} and introduce a potential difference between the two chains. In the bonding-antibonding basis, after bosonization, the potential difference reads
\begin{eqnarray}
V_\perp = -\frac{e V_H}{\pi a_0} \int dx \cos \sqrt{2} \theta_\sigma \cos \sqrt{2} \phi_\sigma,
\end{eqnarray}
and we have to determine $V_H$ such that  $\langle P_H \rangle_{H+V_\perp}=0$. To lowest order in $1/m$, we obtain
\begin{eqnarray}
\langle P_H \rangle &=& \frac{2}{(\pi a_0)^2} \left[\frac{\pi \langle j_\rho \rangle B a}{2 m u_\rho K_\rho} + e V_H\right] \int dx d\tau \left\langle T_\tau [\cos \sqrt{2} \theta_\sigma \cos \sqrt{2} \phi_\sigma](x,\tau)[\cos \sqrt{2} \theta_\sigma \cos \sqrt{2} \phi_\sigma](0,0) \right\rangle =0,
\end{eqnarray}
and in the case of density-density only interactions, $u_\rho K_\rho= u K$. Moreover, according to the non-perturbative definition of the band curvature term \cite{matveev_equilibration_2013}, and as in the bosonic case, we have
\begin{equation}
    R_H=-\frac a {e} \frac{\partial}{\partial \rho_0} [\ln (u_\rho K_\rho)].
\end{equation}

We check in App.~\ref{app:free-fermions} that at low flux, the above relation is valid in the case of a two-leg ladder of non-interacting fermions.

\section{Hall conductance in a free fermion ladder}\label{app:free-fermions}
If we consider a two-leg tight binding ladder with non interacting fermions of Hamiltonian
\begin{eqnarray}
H&=&\sum_{j,p} \left[-t_\parallel(c^\dagger_{j,p} e^{i \left[(-)^p\frac{\Phi}2+\varphi \right] } c_{j+1,p} + \mathrm{H. c.}) -t_\perp c^\dagger_{j,p} c_{j,-p} \right.\nonumber \\
&-&\left. V_\perp (n_{j,1}-n_{j,2}) \right],
\end{eqnarray}
the longitudinal current is
\begin{equation}
j_s = - \frac{\partial H}{\partial \varphi},
\end{equation}
while the Hall imbalance is
\begin{equation}
P_H=-\frac{\partial H}{\partial V_\perp}.
\end{equation}
The Hamiltonian is diagonalized in momentum space, with two bands of fermions having dispersion
\begin{eqnarray}
E_\pm(k)=-2t_\parallel \cos \frac{\Phi} 2 \cos \left( \frac \varphi L ) \right) \nonumber\pm \sqrt{t_\perp^2 + \left[2t_\parallel \sin \frac{\Phi} 2 \sin \left( \frac \varphi L ) \right) -V_\perp\right]}.
\end{eqnarray}
For $V_\perp=0$,$\varphi=\Phi=0$, when $t_\perp < 2t_\parallel \sin(\pi \rho_0/2)$, both bands are partially filled. We will from now on consider a system satisfying such condition and turn on small values of $\varphi,\Phi,V_\perp$. To calculate the Hall resistivity in such system,
we will impose that $\langle P_H\rangle=0$.
When both $V_\perp\ne 0$ and $\Phi\ne 0$, $E_r(k)\ne E_r(-k)$, so the Fermi points are not anymore symmetric around $k=0$. We call $k_{F,r}^{(q)}$ the Fermi point of the $r=\pm$ band of sign $q=\pm$.
We have the condition
\begin{equation}
E_+(k_{F,+}^{(+)}) = E_+(k_{F,+}^{(-)}) =E_-(k_{F,-}^{(+)})=E_-(k_{F,-}^{(-)}),
\end{equation}
leading, to lowest order in $\Phi$ and $V_\perp$, to
\begin{equation}
\frac{k_{F,+}^{(+)}+k_{F,+}^{(-)}} 2 = -\frac{V_\perp}{2t_\perp} \Phi, \quad
\frac{k_{F,-}^{(+)}+k_{F,-}^{(-)}} 2 = \frac{V_\perp}{2t_\perp} \Phi,
\end{equation}
so that $\sum_{r,q} k_{F,r}^{(q)}=2\pi \rho_0$ remains constant.
The ground state energy being
\begin{equation}
E_{GS}=\sum_{r} \int_{k_{F,r}^{(-)}}^{k_{F,r}^{(+)}} E_r(k) \frac{dk}{2\pi},
\end{equation}
applying the Hellmann-Feynman theorem and differentiating with respect to $V_\perp$, we obtain the imbalance
\begin{equation}
\langle P_H \rangle = \sum_{r} \int_{k_{F,r}^{(-)}}^{k_{F,r}^{(+)}} \frac{\partial E_r(k)}{\partial V_\perp} \frac{dk}{2\pi}.
\end{equation}
In the limit of low flux, $\frac{\partial E_r(k)}{\partial V_\perp} \simeq r \frac{V_\perp}{t_\perp}$, so
\begin{equation}
\langle P_H \rangle = \frac{V_\perp}{2\pi t_\perp} (k_{F+}^{(+)}- k_{F+}^{(-)} - k_{F-}^{(+)} + k_{F-}^{(-)}).
\end{equation}
Since the result is already of first order in $V_\perp$, we can take the limit $V_\perp \to 0$ to evaluate the Fermi momenta. Doing so, we find
\begin{equation}
\langle P_H \rangle = -\frac{V_\perp}{\pi t_\parallel \sin (\pi \rho_0 )}.
\end{equation}
So far, we have not considered the contribution of the Aharonov-Bohm flux $\varphi$. In the limit of $V_\perp \to 0$, the Aharonov-Bohm flux gives a contribution to the Hall imbalance $\langle P_H \rangle = \frac{\Phi \cos (\pi \rho_0)}{\pi \sin (\pi \rho_0)} \frac{\varphi}{L}$. These two contributions have to cancel each other, so the Hall imbalance depends on the Aharonov-Bohm flux via $V_\perp = -t_\parallel \cos (\pi \rho_0) \Phi \frac{\varphi}{L}$.
The longitudinal current depends on $\varphi$ as $j_s = \frac{2t_\parallel}{\pi} \sin(\pi \rho_0 ) \frac{\varphi}{L}$,
so we obtain
\begin{equation}\label{eq:hall-noninteracting-ladder}
    V_\perp = -\frac\pi 2 \frac{\Phi}{\tan(\pi \rho_0)} J
\end{equation}
The relation between the longitudinal current and the Aharonov-Bohm flux implies that the charge stiffness of the ladder is $\mathbb{D}=2t_\parallel \sin(\pi \rho_0)/\pi$.
Taking its logarithmic derivative, we recover the coefficient in Eq.~(\ref{eq:hall-noninteracting-ladder}).
So the relation established perturbatively using bosonization is verified in a non-interacting case with non-quadratic dispersion by a direct calculation.

\section{Hall resistance in a Galilean invariant system}\label{app:rh-galil}
In a Galilean invariant system, a simple argument can be given
to explain the dependence of the Hall resistance on the particle density. While it is reminiscent of the argument used in the two-dimensional case \cite{girvin_houches} to show that the Hall resistance is inversely proportional to the density, it does not rely on the Lorentz covariance of the electromagnetic vector potential, so it applies equally to artificial gauge fields. It only requires invariance under Galilean boosts along the chain direction. We will choose a gauge in which the vector potential is along the rungs $A_y=B x $ and $A_x=0$. Under a Galilean boost generated by the time-dependent unitary operator
\begin{eqnarray}\label{eq:unitary-galilean}
U_v &=& e^{i v P t} e^{-i m v X},
\end{eqnarray}
where
\begin{eqnarray}
P=-i \int dx \sum_n \psi^\dagger_n \partial_x \psi_n, \quad X= \int dx x \psi_n^\dagger \psi_n,
\end{eqnarray}
are respectively the total momentum and position operators,
the Hamiltonian becomes
\begin{eqnarray}
U^\dagger_v H U_v - i U^\dagger_v \partial_t U_v= U^\dagger H U + vP,
\end{eqnarray}
while the annihilation operator becomes
\begin{eqnarray}
\psi(x)=e^{i m v (x+vt)} \psi(x+vt).
\end{eqnarray}
the interchain hopping is
\begin{equation}
-t_\perp \int dx (e^{i e B a x} \psi^\dagger_1(x) \psi_2(x) + \mathrm{H. c.}),
\end{equation}
and after the Galilean boost, we find a vector potential $A'_y=B (x-vt)$ in the moving frame, so we have an electric field $E_y=v B$.
Moreover, under the boost, the current $j_x$ is shifted by an amount $e \rho_0 v$, so Galilean boost allow us to map a system in equilibrium in the ground state to a system with a longitudinal current $\langle j_x\rangle=e \rho_0 v$ and a transverse electric field $E_y=v B$. This leads to a Hall resistance $R_H=E_y a/(\langle j_x \rangle B) = a/(e \rho_0)$, as we have found in our perturbative calculation. 

%

\end{cbunit}

\end{document}